\documentclass{aa501}
\usepackage{psfig}
\usepackage{graphicx}

\newcommand{\ie}{{i.e. }}
\newcommand{\etal}{{et~al.}}
\newcommand{\eg}{{e.g.}}
\newcommand{\lsim}{\,\lower2truept\hbox{${<\atop\hbox{\raise4truept\hbox{$\sim$}}}$}\,}
\newcommand{\gsim}{\,\lower2truept\hbox{${>\atop\hbox{\raise4truept\hbox{$\sim$}}}$}\,}

\begin{document}

\title{Removing $1/f$ noise stripes in cosmic microwave background anisotropy observations}

\titlerunning{Removing $1/f$ noise stripes in CMB anisotropy observations}

\author{
    D.~Maino \inst{1} \and C.~Burigana \inst{2} \and K.~M. G\'orski
\inst{3,4}
    \and N.~Mandolesi \inst{2}  \and M.~Bersanelli \inst{5}
        }

\offprints{D.~Maino (maino@ts.astro.it)}

\institute{Osservatorio Astronomico di Trieste, Via G.B.~Tiepolo 11, I-34131, Trieste, Italy
\and Istituto TeSRE, Consiglio Nazionale delle Ricerche, Via Gobetti 101, I-40129, Bologna, Italy
\and ESO, European Southern Observatory, Karl-Schwarzschild Str. 2, D-85748, Garching, Germany
\and Warsaw University Observatory, Aleje Ujazdowskie 4, 00-478, Warsaw, Poland
\and Universit\'a degli studi di Milano, Via Celoria 31, I-20131, Milano, Italy
}

\date{Received; accepted }

    \abstract{
    Removal of systematic effects is crucial in present and future
    CMB experiments mapping large fraction of the sky.
    Accurate CMB measurements ask for multi-feed array instruments
    observing the sky with a redundant scanning strategy covering
    the same sky region on different time scales and with
    different detectors for a better control of systematic
    effects.
    We investigate the capability to suppress $1/f$ noise features
    in Time Ordered Data (TOD) by using the destriping technique described
    in Maino \etal~1999, under realistic assumptions for
    crossing condition between different scan circles and sky signal
    fluctuations on small angular scales.
    We perform, as a working case, {\sc Planck}-LFI
    simulated observations
    with few arcminutes pixel size convolved with LFI beam
    resolutions.
    In the noiseless case for crossing condition based
    on pixels with side larger than the input one, the destriping algorithm
    inserts extra-noise in the final map of the order of $\sim \mu$K in $rms$
    and few $\mu$K in peak-to-peak amplitude at 30~GHz.
    However including instrumental noise (white and $1/f$ noise)
    in the TOD, the impact of the sky signal on the destriping is found
    to be very small. In addition, for crossing condition based on pixels with side half of
    the one of the final map (typically $\sim$ 1/3 of the FWHM), we find only
    a small improvement ($\sim 1$ \% level) in the destriping quality with respect
    to the case when crossings are searched on pixels with same size of the final map
    one. We can conclude that the receiver noise is the driver for destriping quality.
    We extend the analysis to high values of the knee frequency and find that,
    although significantly suppressed by destriping, the residual additional noise
    $rms$ is $\sim 31$\% larger than the pure white noise $rms$ at $f_k = 1$~Hz which
    could be a critical issue in the extraction of CMB angular power spectrum.
    We verified that the approximation of the $1/f$ noise on
    averaged scan circles as a single baseline still works well
    even for these high values of the knee frequency.
    Furthermore, by comparing simulations with different noise
    levels and different sampling rates,
    we find that the destriping quality does not significantly depend on the
    receiver sensitivity whereas it improves proportionally to the improvement of
    sampling rate. Therefore given a noise level, the higher the
    sampling rate, the better the destriping quality.
    This paper is based upon {\sc Planck}-LFI activities.
    }

\maketitle

\bibliographystyle{astron}

\section{Introduction}

In recent years a substantial improvement in measurements of
Cosmic Microwave Background (CMB) radiation has taken place
leading to several CMB anisotropy detections which support the
gravitational instability scenario for structures formation (Smoot
\etal~1992, G\'orski \etal~1996) and a universe with
$\Omega_0\sim1$ ({\bf Miller \etal~1999,} De Bernardis \etal~2000, 
Balbi \etal~2000, Pryke \etal~2001). A
more complete analysis of current data implies multi-dimensional
fits to jointly constrain several cosmological parameters (e.g.
Lineweaver 1998, Tegmark \& Zaldarriaga 2000). These estimations
relay always on the accuracy by which foreground emissions and
experimental systematic effects are known and properly accounted
for. One of the major problems in present and future CMB
anisotropy experiments, are all those possible systematic effects
which affect the experiments in several ways.

In this respect a space mission is the optimum being free from the
unwanted contamination from ground and Earth atmosphere emission.
The Microwave Anisotropy Probe
($MAP$\footnote{http://map.gsfc.nasa.gov/}) satellite (Bennett
\etal~1996) by NASA has begun its mission aimed at all-sky imaging
of the last scattering surface of CMB photons, at several
frequencies and with high angular resolution. In 2007 the {\sc
Planck}
satellite\footnote{http://astro.estec.esa.nl/SA-general/Projects/Planck/}
(Mandolesi \etal~1998, Puget \etal~1998) by ESA, will probe the
very early phase of the Universe with even higher spectral
coverage, sensitivity and angular resolution. Both satellites will
operate around the $L_2$ Lagrangian point of the Sun-Earth system.
This will allow a considerable rejection of the Sun, Earth and
Moon radiation (see, e.g., Burigana \etal~2000), and the adopted
redundant scanning strategy observing the same sky region several
times on different time scales and by different detectors, allows
a high control of systematic effects. In particular, {\sc Planck}
is a third generation of CMB mission, covering the widest
frequency range ($\nu \sim$~30~-~900~GHz) ever probed,
necessary for a high accuracy subtraction of foreground
contamination, and reaching a sensitivity per 0.3$^\circ$ pixel
size of few~$\mu$K at each frequency channel $\nu \lsim 400$~GHz.
Therefore an accurate monitoring and removal of systematic effects
is crucial to reach the planned scientific objectives.

In the context of the {\sc Planck}-LFI (Low Frequency Instrument)
a detailed study of the major sources of systematic effects has
been carried out. Burigana \etal~(1998) and Mandolesi \etal~(2000a)
considered the impact of main beam distortions on CMB
observations. Maino \etal~(1999) took into account the problem of
the so-called $1/f$ noise due to gain fluctuations in HEMT
amplifiers in {\sc Planck}-LFI receivers. A detailed study of the
Galactic straylight contamination in LFI observations has been
carried out by Burigana \etal~(2001), where the effects of optical
distortions are compared with the $1/f$ noise contamination.

The $1/f$ noise typically leads to stripes in the final map
(Janssen \etal~1996) altering the statistical properties of the
cosmic signal which is particularly relevant for CMB anisotropy.
This effect can be parameterized by the knee-frequency $f_k$ (\ie
the frequency at which the white and $1/f$ noise power spectra are
equal) which should be as small as possible with respect to the
spin-frequency of the satellite (\eg~for {\sc Planck} $f_s =
0.0167$~Hz corresponding to 1~r.p.m. spin velocity). It is
therefore crucial to properly correct this effect both by hardware
and software techniques. The LFI, pseudo-correlation receivers,
are properly designed to reduce the $1/f$ noise (Bersanelli
\etal~1995). Delabrouille (1998) has implemented a technique for
destriping {\sc Planck} observations starting from the Time
Ordered Data (hereafter TOD) and taking advantage of the
redundancy of the {\sc Planck} scanning strategy. Maino
\etal~(1999) have considered a similar technique in the more
specific context of {\sc Planck}-LFI observations and using the
current theoretical predictions on LFI receiver properties.
The results show that the destriping
quality is remarkably good, except for the degenerate case in
which all the crossings between different scan circles are
concentrated very close to the ecliptic poles.

Although destriping algorithms are sometimes considered
approximations of proper map making algorithms (recently
implemented for large data time ordered data by, e.g., Natoli
\etal~2001 and Dore \etal~2001) they should be considered as
methods to remove drifts in the TOD's and returning TODs cleaned
from many classes of systematic effects
(see, e.g., Mennella \etal~2001 for an analysis of periodic fluctuations).
This is more suitable
for many data analysis purposes directly on TODs such as
monitoring of variable and transient sources (Terenzi \etal~2001).
In the following
analysis we make use of maps obtained by co-adding the TOD's
cleaned by destriping codes: this way of proceed is one of the
possible methods to quantify the quality
of the destriping algorithms.

A key issue of destriping techniques is the operative definition of
crossing between two different scan circles, since these techniques
are based on the condition that the observed temperatures in the sky
have to be the same for identical sky directions, although
the samplings are taken along different scan circles at different times.
Of course one can be more or less restrictive on the definition
of the crossing condition: a more restrictive definition
therefore may reduce the number of crossing possibly affecting
the destriping quality while a less one may insert an extra-noise
due to the different sky signals in the two not exactly
coincident sampling directions.

Another issue is the validity of the $1/f$ noise drift approximation
in terms of a constant baseline (offset) for each averaged scan circles
(or significant fraction of it), which in general holds
as long as $f_k$ is not far larger than $f_s$.
Of course, the values of $f_k$ appropriate to LFI receivers
will be probed in future real hardware analysis.
It is therefore
interesting to assess the maximum values of $f_k$ for which the
destriping algorithm still works well removing $1/f$ noise stripes
at a level which does not compromise the determination of CMB
angular power spectrum.

In this paper we want to address these issues, evaluate the impact
of different crossing conditions for different models of the
microwave sky, including instrumental noise. Furthermore we push
the knee-frequency to extremely high, and hopefully unrealistic,
values ($\sim 1$~Hz) to evaluate the destriping quality also for
pessimistic cases.

We want to stress the fact that, although applied for the
specific case of the {\sc Planck} mission, these arguments are
relevant to almost any CMB anisotropy observations. In typical CMB
experiments,the scanning strategy implies repeated observations
of the same sky regions on different time scales such as, \eg,
BEAST (Seiffert 1996), COSMOSOMAS (Watson 1997). Note that the
actual implementation of our destriping algorithm is completely
independent of the details of the scanning strategy and of the
pixelisation scheme\footnote{Our first implementation (Burigana
\etal~1997b) worked in fact with the QUAD-CUBE scheme and the
subsequent applications (Maino \etal~1999) with the HEALPix
scheme (G\'orski \etal~1998) did not imply any modification at
all.}.

This paper is organized as follows. We briefly describe the {\sc
Planck} scanning strategy and the generation of the TOD's in
Sect.~\ref{scanning}. The basic recipes of our destriping
technique and the discussion of the crossing conditions are
presented in Sect.~\ref{destri}. We report our simulation results
in Sect.~\ref{simul}, where we assess the impact of the choice of
crossing condition and intrinsic sky fluctuations on destriping
quality. We dedicate Sect.~\ref{knee} to the evaluation of the
impact of possible high values of the knee-frequency on the
destriping quality. We discuss there also the dependence of the
destriping efficiency on the instrumental parameters (sensitivity
and sampling rate) at different frequencies and introduce the
functional form for noise power spectrum after destriping.
Finally, we discuss our results and draw our conclusions in
Sect.~\ref{conclu}.

\section{{\sc Planck} observations}
\label{scanning}

\subsection{Scanning strategy}

The orbit selected for {\sc Planck} satellite will be a tight
Lissajous orbit around the $L_2$
Lagrangian point of the Sun-Earth system
(Mandolesi \etal~1998, Puget \etal~1998).
The spacecraft spins at
$\sim$ 1 r.p.m. ($f_s \sim 0.0167$~Hz) and, in the simplest scanning
strategy, the spin axis is essentially kept on the antisolar direction
at constant solar aspect angle by a re-pointing of $2.5'$ every hour\footnote{This
simple scanning strategy provides also a smooth and quite uniform sky coverage which
covers $\gsim 99\%$ of the full sky}.
The LFI and HFI (High Frequency Instrument) share the focal plane of an
Aplanatic telescope (see, e.g., Mandolesi \etal~2000b for a discussion
on the advantages of this configuration)
of 1.5 meter size which field of view is at an angle
$\alpha =85^\circ$ from the spin-axis.
Therefore {\sc Planck}
will trace large circles in the sky. In the nominal 14
mission months $\sim 10200$ spin-axis positions will be exploited
covering twice nearly the whole sky, some regions of which
will be covered three times. The actual sampling
frequency (\ie the frequency at which the
sky signal is sampled along a given scan circle) adopted for LFI
is about 3 samplings per FWHM resulting in a different number of
samplings at the four LFI frequencies.

The details of the scanning strategy are not yet frozen and it may or not include
a precession of the spacecraft spin axis about another axis kept
along the antisolar direction re-pointed of $2.5'$ every hour.

\subsection{From TOD to sky maps}
\label{fs}

We have implemented a code (``Flight Simulator'',
Burigana \etal~1997b,~1998 and Maino \etal~1999)
simulating the {\sc Planck} scanning strategy and observations
and applied to the study of some systematic effects.
The relevant geometrical inputs of the code
are the angle $\alpha$ between pointing and spin axis,
beam position on the focal plane as well as beam response function.
Other inputs are parameters describing the noise properties
of the receivers.

We adopt here the nominal
{\sc Planck} scanning strategy, \ie with
14 months of mission and re-pointing of the spin-axis every hour
by $2.5'$ assuming 3 samplings per FWHM, and refer in this section
to the case of simulations at 30~GHz where the nominal resolution
(FWHM) is $33'$.
The generation of instrumental noise is performed directly in
Fourier space and FFT transformed back in real space
(see Maino \etal~1999 for details).

We use the HEALPix pixelisation scheme (G\'orski \etal~1998),
which is the adopted baseline for {\sc Planck} products.
The output of our Flight Simulator code are 4 matrices: {\bf N}, with
pointing pixel number at specified resolution
(usually larger than the beam size for a proper sampling of beam
resolution), {\bf T}, with sky
temperature plus full noise (white + $1/f$ noise), {\bf W}, with sky
signal plus white noise only and {\bf G}, with sky signal only.
Each matrix has $n_s$ rows equal to the number of spin axis
positions ($n_s \sim 10200$) and $n_p$ columns equal to the number of
samplings, weakly dependent on $\alpha$, along a scan circle
($n_p \sim 1980$ at 30~GHz).
The matrices {\bf W} and {\bf G} are computed to evaluate respectively
the
degradation of $1/f$ with respect ideal white noise case and the
impact of the scanning strategy geometry on the observed signal.

From these TOD we build sky maps: making use of {\bf N},
possibly degraded at the desired resolution of the final map,
and {\bf T} we simply coadd the temperatures of those pixels which are
observed several times during the mission. In the same way we
build maps with only white noise, from {\bf W}, and, using {\bf G},
without receiver noise.


\section{Definition of the problem}
\label{destri}

As first pointed out by Janssen \etal~(1996) the effect
of $1/f$ noise can be seen as one additive level
(baseline or offset) different for each scan circle
(or significant fraction of it). Since we work
with scan circles averaged over 1-hour period, we strongly reduce
drifts within the circle and we are left with the ``mean'' $1/f$
noise level for that observing period. Furthermore average is like
a low-pass filter operation and, as long as the $f_k$ is not far
larger than $f_s$, this leaves only the very low frequencies
components of the $1/f$ noise. Therefore it is a
good approximation to model the effect of $1/f$ noise
on the averaged scan circles as a single baseline $A_i$.
We want to recover these baselines in order to
properly adjust the signal.

\subsection{Destriping technique}

The destriping technique considered here, developed by Burigana
\etal~(1997b) and by Maino \etal~(1999), has been derived from the
COBRAS/SAMBA Phase A study
proposal (Bersanelli \etal~1996) and from a re-analysis of Delabrouille
(1998).

The first logical step of this method is the search of the crossings,
i.e. to find the common pixels
observed from different scan circles. This can be done at
the desired resolution, not necessarily the same of the final map,
if poorer than that at which the pixel stream is computed and
stored.
Let us indicate with $N_{il}$, $T_{il}$ and $E_{il}$ the pixel number,
the observed signal and the ``white'' noise level for the pixel in the
$i^{\rm th}$ row (scan circle) and $j^{\rm th}$ column (sampling
along the circle) in matrices {\bf N}, {\bf T} and {\bf (W-G)}
respectively.
Let also be $\pi$ the index that identifies a generic pair of
observations
of the same pixel: of course the index $\pi$ is related to two elements
in the matrix
{\bf N}: $\pi \rightarrow (il,jm)$ where $i$ and $j$ are the
indexes for two different scan circles and $l$ and $m$ the
respective position of the observed pixel.

Following Maino \etal~(1999) baselines are recovered solving the
linear system:
\begin{equation}
\label{1system}
\sum_{\pi=1}^{n_c}\left[
\frac{[(A_i-A_j)-(T_{il}-T_{jm})]\cdot[\delta_{ik}- \delta_{jk}]}
{E_{il}^2-E_{jm}^2}\right]_\pi = 0
\end{equation}
for all the $k=1,...,n_s$ (here $\delta$ is the usual Kronecker
symbol). This simply translates into a set of $n_s$ linear
equation:
\begin{equation}
\sum_{h=1}^{n_s} C_{kh}A_h = B_k,\ \ k=1,...,n_s
\label{system}
\end{equation}
which can be easily solved: the matrix of the coefficients $C_{kh}$ is
positive defined, symmetric and non singular. This last property
is verified provided that there are enough intersections between
different scan circles.
A detailed discussion on the solving of system in
Eq.(\ref{system}) and RAM requirements can be found in
Burigana \etal~(1997b).

\subsection{Crossing condition and sky map resolution}

It is clear that the capabilities of the destriping technique to
recover circles baselines strongly depend on the distribution of
crossing points and on the quantity of crossings between
different scan circles. Of course, these properties are
correlated to each other and depend on the scan angle $\alpha$,
the beam position on the telescope field of view, the selected
scanning strategy and, finally, on the definition of the crossing
condition. This concept is parameterized in our code by the size
of the pixel in which two samplings on two different scan circles
have to fall to be considered as a ``crossing''. Maino
\etal~(1999) demonstrated that for $\alpha = 90^\circ$ an
off-axis beam (typical LFI position with $(\theta_B,\phi_B) =
(2.8^\circ,45^\circ)$ in the focal plane) destripes better than
an on-axis beam ($\theta_B = 0^\circ$ or $\phi_B = 0^\circ $ or
$\phi_B = 180^\circ$) since for the former the crossing points
are spread on a large region around the Ecliptic polar caps
while, for the latter, they are practically only at the poles
causing the system in Eq.(\ref{system}) to be almost degenerate.

\begin{figure}[here]
\centerline{
    \psfig{file=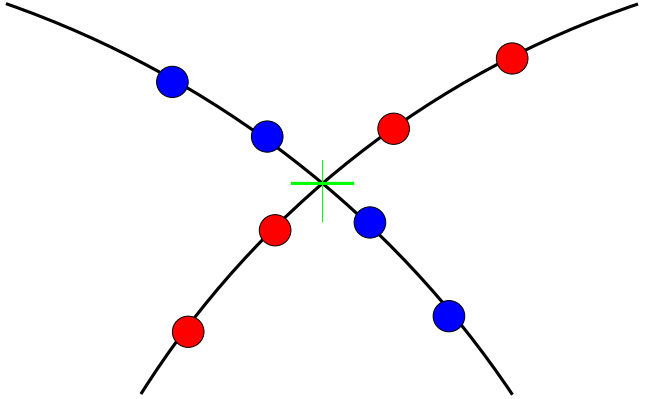,width=8.5cm}
}
\caption{Schematic representation of the crossing between two scan
circles: filled circles are samplings along the circles.}
\label{xing}
\end{figure}

In order to better clarify
the impact of the crossing condition on the destriping quality,
let us consider the situation depicted in Fig.~\ref{xing}: here
filled circles represent samples along two crossing scan circles
while the cross indicates the crossing point. {\bf This could be
regarded as a typical situation in which no sample could be
associated to the crossing point.}

Therefore one possibility would be to interpolate the signals (and
consequently the noise) of the two
samplings to the crossing points and feed the result into
Eq.~(\ref{system}) as in \cite{revenu}. We prefer not to interpolate
but address the problem in a different way {\bf although our approach is formally
equivalent to interpolation if only CMB is considered. However when considering
foreground emission from point sources, that have large power on small angular
scales, interpolation is no more practicable and may lead to large errors.}
We identify pointings by means of HEALPix pixel numbers at a suitable
resolution and two pointings are considered coincident if they have
the same pixel identification number.
This however leaves an uncertainty: pointings can fall into the same pixel
but could not be exactly coincident. This makes the signal
of the two samplings different due to small scales fluctuations
of sky signal due to CMB and foreground emissions.

\begin{figure}[here]
\centerline{
    \psfig{file=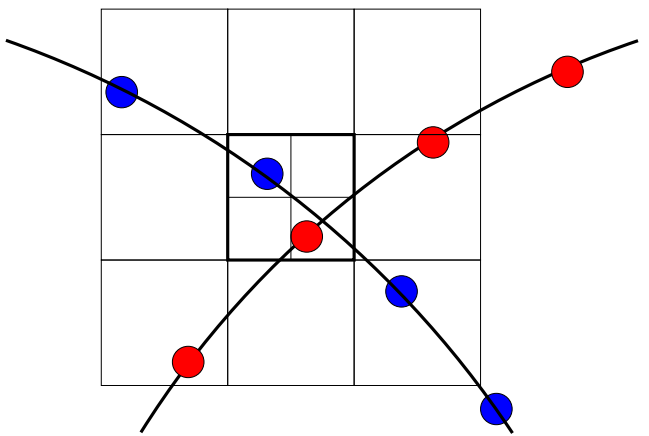,width=8.5cm}
}
\caption{Two samplings falling in the same crossing pixel (the thick
one):
the ``true'' pointings instead fall into different input map
pixels with different sky signal that introduces an ``extra'' noise.}
\label{pointing}
\end{figure}

Since the sky signal has power also on small angular scales
(e.g., due to discrete point sources) it is crucial to have a sky
pixelised with the highest
possible resolution in order to follow these fluctuations. It is
also important to have a crossing pixel dimension small enough to
be as close as possible to the real samplings along the scan
circles.
The ideal case would be having a map with super-resolution and
performing the crossing search at a resolution closer to the
angular dimension of a single sampling. Only in this case is it
possible to evaluate the true capabilities of a destriping
algorithm. However it is not necessary to go to very high
resolution ($\lsim 1'$) since, at least for LFI, the highest
resolution is $10'$ at 100~GHz, which means $\simeq 3.3'$ per
samplings.

Reducing the crossing pixel size has the effect of dramatically
lowering the number of crossings between scan circles, possibly
affecting the destriping quality.
It is however worth to note that the crossings are
$\gsim 10^7$ (on a pixel $\gsim 3.5'$, see, e.g., Table~\ref{table1})
while baselines to be
determined are at least 3 orders of magnitude less.
In the next section we will present accurate simulations that quantify
the impact on the destriping
quality of sky signal fluctuations on small
angular scales and crossing accuracy between scan circles.

\section{Simulations}
\label{simul}

We consider input maps with $\mathcal{N}_{\rm side}= 1024 $
corresponding to $3.5'$ pixel size. We use the baseline for the
{\sc Planck} scanning strategy with $\alpha=85^\circ$ and the
spin-axis always on the Ecliptic plane. Beam position in the
focal plane is chosen to be a worst case with respect the
distribution of crossings between scan circles: crossings are
limited to a small region around the polar caps. Within the
actual focal plane arrangement this results in a position
$(\theta_B,\phi_B) \sim (4^\circ,180^\circ)$.

Our sky model is composed by primordial CMB anisotropy
according to a standard CDM model\footnote{$\Omega_{CDM}=0.95, \Omega_b
= 0.05$ and $H_0 = 50$~Km/s/Mpc} normalized to $COBE$, plus a model for the
Galactic emission.
This model has the spatial template of Galactic dust emission as in
Schlegel \etal~(1998) which has high angular resolution and
therefore includes signal fluctuations on small scales.
The emission of this map has been normalized to take into account
dust, free-free and synchrotron emission at 30~GHz as derived from
$COBE$-DMR data (Kogut \etal~1996).
A realization of extra-galactic point sources has been derived from
number source counts extracted from a
Poissonian distribution (Toffolatti \etal~1998).

\subsection{Pure sky signal}
\label{puresky}
\begin{figure*}[t]
\centerline{
    \psfig{figure=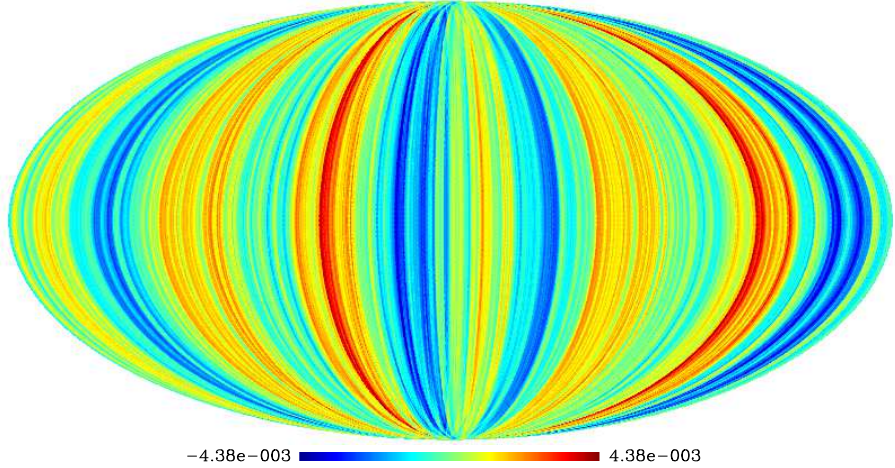,width=12cm,angle=90}
} \caption{Residual stripes after subtraction of the ``destriped''
map with pure sky signal and the input map for a model including
extra-galactic sources and using a crossing pixel dimension of
$13.7'$. Units are mK at 30~GHz. Note that the effect in principle
could be relevant being around 8$\mu$K in peak-to-peak amplitude.}
\label{stripesfig}
\end{figure*}
\begin{table*}
\caption{Peak-to-peak amplitude and $rms$ of the REN as
a function of the crossing pixel size for CMB,CMB + Galaxy and CMB +
Galaxy + Extra-galactic sources.}
\label{table1}
  \begin{center}
    \begin{tabular}{lccc}
    \multicolumn{4}{c}{CMB} \\
    \hline
    Xing pixel & \# pairs & $rms$ & peak-to-peak \\
    \hline
    \ $6.8'$  & $\sim 2\times 10^7$ & 0.386$\mu$K & 2.57$\mu$K \\
    \ $13.7'$ & $\sim 7\times 10^8$ & 0.959$\mu$K & 5.97$\mu$K \\
    \hline
    & & & \\
    \multicolumn{4}{c}{CMB + Galaxy} \\
    \hline
    Xing pixel & \# pairs & $rms$ & peak-to-peak \\
    \hline
    \ $6.8'$  & $\sim 2\times 10^7$ & 0.387$\mu$K & 2.58$\mu$K \\
    \ $13.7'$ & $\sim 7\times 10^8$ & 0.960$\mu$K & 5.97$\mu$K \\
    \hline
    & & & \\
    \multicolumn{4}{c}{CMB + Galaxy + Extra-galactic Sources} \\
    \hline
    Xing pixel & \# pairs & $rms$ & peak-to-peak \\
    \hline
    \ $6.8'$  & $\sim 2\times 10^7$ & 0.615$\mu$K & 4.16$\mu$K \\
    \ $13.7'$ & $\sim 7\times 10^8$ & 1.408$\mu$K & 8.62$\mu$K \\
    \hline
    & & & \\
    \multicolumn{4}{c}{CMB + Galaxy + Extra-galactic Sources + Noise} \\
    \hline
    Xing pixel & \# pairs & $rms$ & peak-to-peak \\
    \hline
    \ $6.8'$  & $\sim 2\times 10^7$ & 31.884$\mu$K & 286.37$\mu$K \\
    \ $13.7'$ & $\sim 7\times 10^8$ & 31.881$\mu$K & 288.39$\mu$K  \\
    \hline
    & & & \\
    \multicolumn{4}{c}{Pure noise} \\
    \hline
    Xing pixel & \# pairs & $rms$ & peak-to-peak \\
    \hline
    \ $6.8'$  & $\sim 2\times 10^7$ & 31.889$\mu$K & 286.11$\mu$K \\
    \ $13.7'$ & $\sim 7\times 10^8$ & 31.871$\mu$K & 285.78$\mu$K \\
    \hline
    \end{tabular}
  \end{center}
\end{table*}

We want to evaluate the impact of intrinsic sky signal
distribution on the destriping. To quantify the effect, we first
consider simulations with no instrumental noise. The search for
common pixels between different scan circles is based upon pixels
with size $\geq 3.5'$. The output map has pixel size of $13.7'$
which is close to the Nyquist frequency for the 30~GHz beam FWHM.
After destriping we subtract the ``destriped'' map from the map
obtained simply co-adding pixels according to the scanning
strategy. As we expect we found a small excess of noise since
pixels that are considered coincident by the destriping
algorithm, have indeed different sky signals (\eg~see
Fig.~\ref{pointing}). The destriping algorithm tries to adjust
this signal difference as if it was due to different offsets in
the scan circles. In Fig.~\ref{stripesfig} we show the residual
extra noise (hereafter REN) for the sky model which includes
extra-galactic point sources: stripes are clearly inserted by the
destriping code.

In Table~\ref{table1} we report $rms$ and peak-to-peak amplitude
of the REN as a function
of the dimension of pixel used for searching for common
observations between scan circles.
The input signal includes a pure CMB sky, CMB plus Galactic signal
and CMB plus Galactic emission and extra-galactic radio sources.
The results at 30~GHz are clearly the worst case for LFI
with respect to the contribution of foreground emissions.

Of course, when the crossing pixel has the same size of pixel input
map, the effect is zero as expected and therefore we do not report
this results. The REN increases, as expected when enlarging the
crossing pixel size, particularly for the case with
extra-galactic point sources due to their power on very small
angular scales. The situation with pure CMB sky is less dramatic
due to the smooth sky gradient.

Note that the $rms$ and peak-to-peak amplitude of the
effect lie in the range $0.4 - 1.4\mu$K and $\sim 6.0 - 8.6\mu$K
respectively on a pixel of $13.7'$. These numbers have to be
compared with the expected sensitivity at 30~GHz re-scaled to
$13.7'$ pixel size, which is of the order of 15$\mu$K.
It is important to remind that the goal for the {\em
overall} budget for systematic effects is few $\mu$K
and this effect could contribute in principle in non negligible way
to the budget.

\subsection{Sky signal and full noise}

We include in our simulations the effect of instrumental white
and $1/f$ noise with a knee frequency of about 0.1~Hz. The sky
signal include CMB, Galactic emission and extra-galactic radio
sources (this is close to a realistic scenario).

To assess the impact of sky signal fluctuations and crossing
condition on the destriping efficiency we compare two sets of
simulations: the first with pure noise and the second with noise
plus signal. In both cases, as in Sect.~\ref{puresky}, the output
map has $13.7'$ pixels while searching for crossing is performed
at $13.7'$ and $6.8'$.
Any difference between the two cases indicates the impact
of sky signal and crossing condition on destriping quality.

Results show that pure noise case is almost identical to the
case which includes sky signal: differences in REN are
of the order of 0.13\% and 0.56\% when crossings are searched on
6.8$'$ and 13.7$'$ pixels sizes respectively. Furthermore we
find very little improvement in destriping quality when
crossings are based on pixel sizes smaller that the one of the
final map (typical differences are below 0.5\%).

This means that the quantity which
determines the destriping quality is the noise component while the
sky signal plays a minor role adding only small degradation.

A crucial estimator of the $1/f$ stripes impact on CMB anisotropy
measurements is its angular power spectrum. It can be directly compared
to the CMB angular power spectrum and with the spectra that
characterize other classes of both instrumental and astrophysical contaminations.

In the case of instrumental pure white noise and
in the limit of full sky coverage
the angular power spectrum is given
by
\begin{equation}
C_\ell \simeq \frac{4\pi}{N_{\rm pix}^2}\sum_{i=1}^{N_{\rm pix}}\sigma_i^2\,
,
\label{cl}
\end{equation}
where $\sigma_i$ is the noise in the $i^{\rm th}$ pixel and
$N_{\rm pix}$ is the total number of pixels.

\begin{figure}[t]
\centerline{
    \psfig{figure=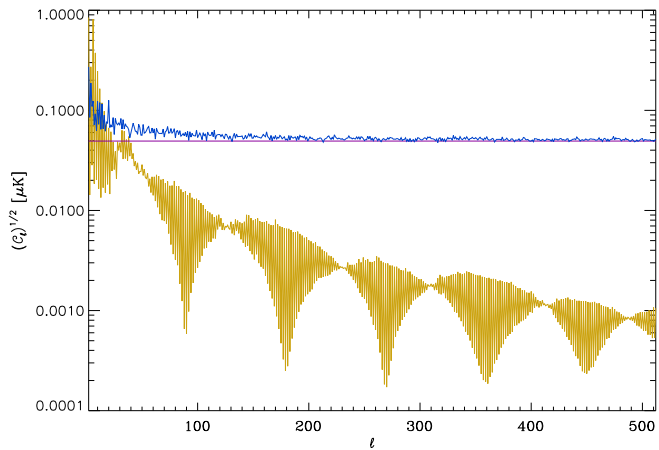,width=9cm}
}
\caption{Angular power spectrum of the REN
for the pure signal case (CMB + Galaxy + Extra-galactic sources; grey line),
compared with the level of the white noise power spectrum (straight solid line)
and the noise spectrum after destriping (pure noise case - solid line).}
\label{diff}
\end{figure}

In Fig.~\ref{diff} the angular power spectrum of the
REN for the pure signal case (CMB + Galaxy + Extra-galactic sources)
is shown compared with the expected level of white noise and the noise spectrum after
destriping (pure noise case).
It is clear that most of the excess at low $\ell$s is due to
residual $1/f$ noise after destriping (see Maino \etal~1999). For higher
$\ell$ ($\gsim 50$),
where the destriping algorithm works better and the noise spectrum is almost
white, the contamination is below the white noise
level.
This again underlines the small impact of signal correlation in
the destriping algorithm and, furthermore, leads to the conclusion
that, in this realistic case of sky
signal and instrumental noise, the destriping algorithm does not
introduce extra-correlation since the residual after
subtraction of the input sky is almost white noise plus a
residual $1/f$.

\section{Destriping quality versus knee-frequency}
\label{knee}

Since the sky signal contributes only marginally to the destriping
quality when instrumental noise is included, we concentrate our
attention on instrumental noise only with increasing values of the
knee-frequency. We proceed with the analysis of high values of
the knee-frequency, up to 1~Hz. This in principle might impact
the destriping quality because the filter applied to the data
(i.e. the mean over 1-hour period) could not be adequate and so
the constant baseline approximation of residual noise

\subsection{Simulations at 30~GHz}

We explore the intervals of $f_k$ ranging from 0.025~Hz up to
1~Hz: the former being derived from theoretical computation as in
Burigana \etal~(1997a) and Seiffert \etal~(2001), the latter being a pessimistic
conservative case. These knee-frequencies span from values
comparable with the spin frequency to values $\sim 100$ times
larger. Note that for LFI the baseline requirement is set to $f_k
< 0.05$~Hz. The output map has the usual $13.7'$ resolution and
the crossing pixel size is $13.7'$ following the results of
previous section.

In Table~\ref{kneerange} we report results of the fractional,
compared to the white noise, REN $rms$ before and after
destriping, together with their ratio. It is clear the increasing
degradation for increasing $f_k$ after destriping: this reaches
value of 13\% to 30\% of the white noise $rms$ at 0.4 and 1~Hz
respectively.
\begin{table}
\caption{Amplitude of the fractional REN before and after destriping
and the corresponding ratio as a function of the knee-frequency.}
\label{kneerange}
    \begin{center}
        \begin{tabular}{lccc}
        \hline
        \ $f_k$ & Before & After & Ratio\\
        \ [Hz]  & [\%]   & [\%]  &  \\
        \hline
        \ 0.025 &  63.6  &  1.13 & 56.3 \\
        \ 0.1   & 177.8  &  3.71 & 47.9 \\
        \ 0.4   & 427.9  &  13.4 & 31.9 \\
        \ 1.0   & 725.7  &  30.7 & 23.6 \\
        \hline
        \end{tabular}
    \end{center}
\end{table}
The ratio is another indicator of destriping degradation: it
decreases and will approach unity for extreme high values of
$f_k$. It is however interesting to note the small range of values
exploited by the ratio compared to the wide range in $f_k$
indicating a good capability of destriping to control the residual
noise.

\begin{figure*}[ht]
\centerline{
    \psfig{figure=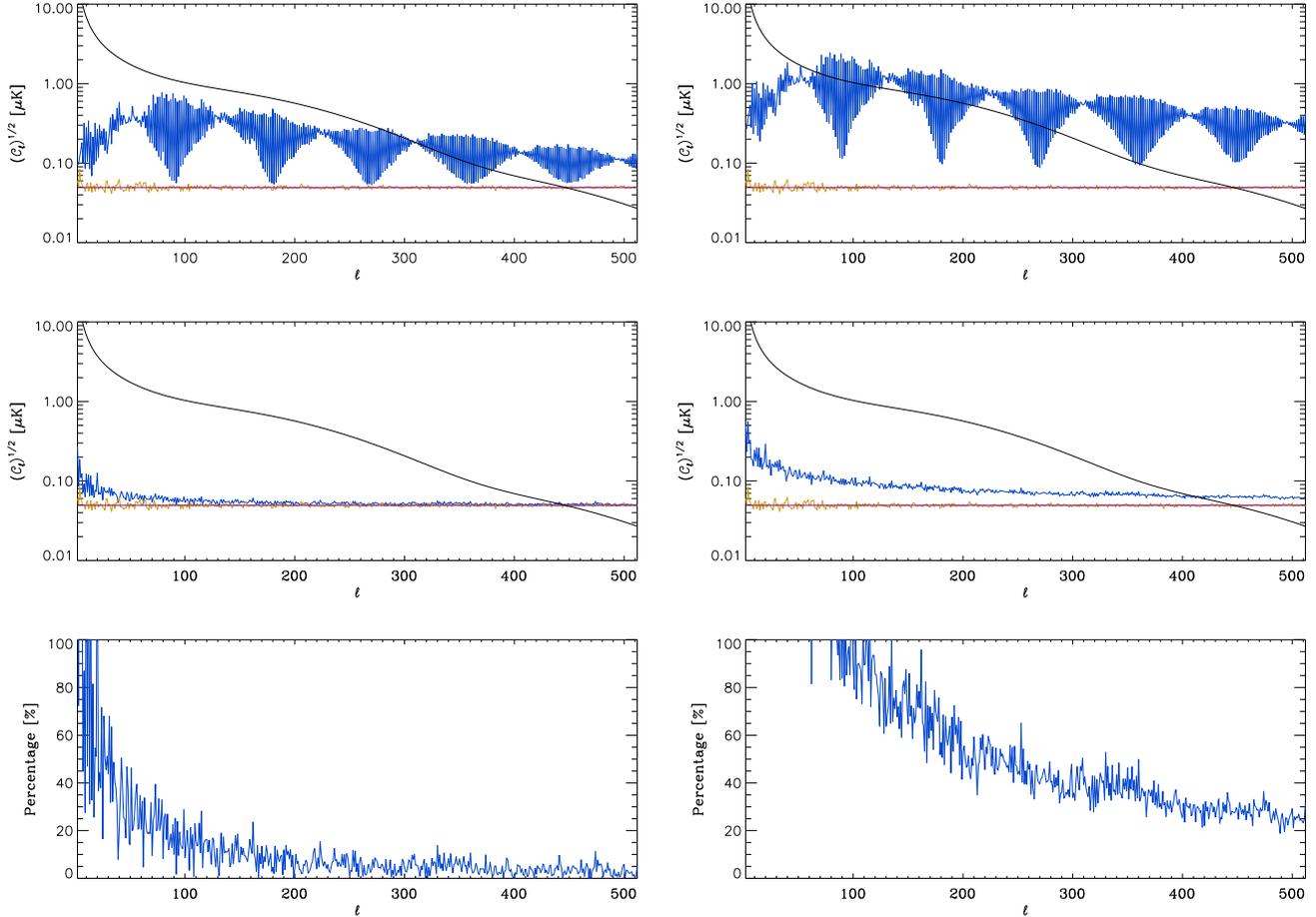}
} \caption{Angular power spectrum before (top) and after (middle)
destriping for $f_k = 0.1$~Hz (left) and for $f_k=1$~Hz (right)
compared to the theoretical level of the white noise and the CMB
power spectrum convolved with the nominal FWHM at 30~GHz. In the
bottom panel the residual noise in percentage of white noise level
is reported as a function of multipole.} \label{ps}
\end{figure*}

The increase of the degradation for increasing knee-frequency
would have an impact on the angular power spectrum obtained
inverting the sky map after destriping. In Fig.~\ref{ps} we
report the angular power spectra before and after destriping for
$f_k = 0.1$~Hz (left) and $f_k = 1$~Hz (right) respectively. The
height of the blobs increases as well as the low-multipole tail
of the residual noise. This of course could be a problem in the
extraction of the CMB angular power spectrum in particular if the
structures related to low-$\ell$ tail show a non-gaussian
distribution. This can be verified by Wiener filtering residual noise map
taking the low-$\ell$ tail as a signal and subtract from it the white noise level
derived from the high multipoles. It can be shown that only for very
low $f_k$ the distribution of the residual noise could be approximated as a
gaussian, while for larger $f_k$ stripes appear: this is indeed the level of
stripes that the destriping algorithm is not able to remove since it
is completely embedded into white noise. However this effect has to carefully
considered in CMB data analysis in presence of $1/f$ noise.

We also consider simulations at 30~GHz with $f_k = 0.4$ and 1~Hz in
which we try to recover two baselines per circle instead
of a single baseline. After destriping the level of extra noise is
around $\sim 14\%$ and $\sim 31\%$, respectively for $f_k = 0.4$
and 1~Hz almost the same as in the single baseline case.
This means that despite the fact the $f_k \simeq 60 \times f_s$,
a single baseline is a good approximation of the averaged
$1/f$ noise.

\subsection{Dependence on the sampling per circle: simulations at 100~GHz}

To complete our estimates on the dependence of the destriping efficiency
and of the residual $1/f$ noise on the instrumental parameters
at different frequencies we extended our simulations to
include a case with a
larger number of samplings per scan circle.

We performed simulations, including only instrumental noise,
at LFI 100~GHz channel that has
a nominal angular resolution of 10$'$ and a noise level per unit
integration time which is about 3 - 4 times larger than
a single channel at 30~GHz. We exploit the same geometrical configuration of the
previous 30~GHz case and $f_k = 0.4$~Hz. Output
maps are at $\mathcal{N}_{\rm side} = 1024$ with 3.5$'$ pixel size
that is appropriate for the 10$'$ angular resolution. Crossings
between scan circles are searched for on 3.5$'$ and 1.75$'$ pixel
size. For both cases we obtain the same results: the fractional REN
after destriping is $\sim$ 4.3\% of the white noise.

For further checking the dependence of destriping quality upon white noise
level and number of samples along a scan circle, we carried out
another simulation with the same parameters as before but
with the white noise level properly scaled in order to have the
same noise level per unit integration time as at 30~GHz.
We obtain a fractional REN after destriping of $\sim 4.4\%$ of the
white noise level, almost equal to the previous result.
Comparison with results in Section~5.1
indicates that, for the same value of the knee-frequency and white noise
level, the larger the number of samplings along a scan circle, the better
destriping quality is achieved.

\subsection{Upper limits on the $1/f$ noise impact}

As shown above, the residual $1/f$ noise contamination after the destriping,
expressed in terms of fractional additional noise with respect the pure white noise,
significantly decreases with increasing sampling rate.

In our simulations we have located the beams in a worst-case position
regarding the destriping quality assuming an angle $\alpha =
85^\circ$ and a scanning strategy without precession of the
spin-axis. The inclusion of the precession although affecting the
distribution of integration time per pixel, might partially
increase the destriping performance (Maino \etal~1999).

In addition Burigana \etal~(2001)
have shown that
the quality of the destriping algorithm is not significantly
affected by the main beam distortion effect and by the beam rotation.
The interpretation of this result is analogous to that
discussed in Sect.~4 to explain the results of Table~1: differences
of few $\mu$K between the two signals (without noise) of a crossing
used in the destriping algorithm, due either to the underlying
different sky or to different beam orientations,
are overwhelmed by the instrumental noise fluctuations.

These considerations strongly indicate
that the results reported in Table~2,
strictly holding for the configuration considered there,
can be regarded as more general upper limits
applying to the residual $1/f$ noise contamination
after destriping.
We can then exploit the results of Table~2 to
derive analytical upper limits for the extra noise
after destriping
with respect to the case of pure white noise
as a function of the knee-frequency.
Fitting the results we obtain a power law of the form:

\begin{equation}
{\rm log}(rms) \lsim 0.898 \times {\rm log}(f_k) - 0.517\ ,
\label{analy}
\end{equation}
where $rms$ is expressed in terms of white noise level (\%) and
$f_k$ is in Hz. The accuracy in $rms$ of Eq.~(\ref{analy}) is
$\lsim 4$\% in the explored range of $f_k$.

\section{Discussion and conclusions}
\label{conclu}

We have reported here a detailed study on removal of $1/f$ noise features
in CMB observations directly on TODs to obtain cleaned TODs. For this
purpose we have made use of the
destriping algorithm
developed by Burigana \etal~(1997b) and Maino \etal~(1999) taking into
account the impact of intrinsic sky signal fluctuations on the
angular scale used to recover common observations between
different scan circles and by extending
the range of the most relevant instrumental parameters, such as
sensitivity, knee frequency and sampling rate.

We have verified that, with pure sky signal in absence of
noise, relaxing the crossing condition
has the effect of
introducing REN which appears
as stripes in the map. This is
particularly
important when including in simulated input sky the contribution
coming from radio sources.
The typical $rms$ effect is of the order of 0.6$\mu$K and shows
peak-to-peak amplitude of $7\mu$K at 30~GHz.
This effect, although small, may be in principle important since the overall
budget for systematic effects for LFI is of few $\mu$K.
On the other hand, we find that this effect is practically {\bf eliminated}
when the instrumental noise is taken into account,
since the noise component is the dominant one.

We explore a wide range of possible values of
knee-frequency from 0.01~Hz up to 1~Hz. The extra noise after
destriping increases with increasing values of $f_k$ and we derive
an analytical expression for the upper limits of the residual
$1/f$ noise contamination in terms of fractional excess of noise
with respect to the pure white noise case.
In the worst case the residual contamination
may be relevant if the knee frequency is not strongly reduced
via hardware; for example,
for $f_k = 0.4$~Hz (1~Hz) we find a fractional REN of about 13\% (30\%).

We have verified that a single baseline approximation in the use of the
destriping code is
still valid even for high values of the knee frequency,
since up to $f_k = 1$~Hz no significant improvement is found by using
two baselines per scan circle.

Furthermore we find that the
destriping quality is only very slightly improved when crossings
are searched on pixels smaller than the ones of the final map.

Finally, our simulations at 100~GHz
show that the destriping quality does not depend on the intrinsic
white noise level,
whereas it improves about proportionally to the number
of samplings along the scan circle.
Analogously, from the comparison with the results obtained
in a previous simulation work (Maino \etal~1999),
we find that the current {\sc Planck} 2.5$'$ shift in spin axis
re-pointing, better than the 5$'$ shift of the Phase A study,
provides a better sampling along the azimuthal direction
which further improves the destriping quality.

This extensive study of the destriping algorithm capabilities and
dependencies upon different configurations confirms the robustness
of the technique and allows to improve its use.
Furthermore the general approach of this technique,
allow its application to most of the present and future CMB experiments
where a redundant scanning strategy is considered.
In CMB observations
the complexity and the mixing
of many classes of instrumental and astrophysical systematic
effects requires a particular care.
In fact new detailed analyses
based on both measured hardware performances provided by laboratories
tests and on a more detailed description of instrument and mission
parameters (including
focal plane assembly, scanning strategy, detailed knowledge of
sampling strategies) will be carried out in the future.


\begin{acknowledgements}

It is a pleasure to thank F.~Arg\"ueso G\'omez and L.~Toffolatti for the
point source map. The use of the HEALPix package is acknowledge.
We would like to thank Uro\v{s} Seljak and Matias Zaldarriaga for
making their CMBFast code publicly available. We are indebt with
M.~Maltoni, M.~Malaspina, L.~Danese, A.J.~Banday and B.D.~Wandelt
for useful discussions. DM and CB warmly thank the TAC in
Copenhagen for hospitality in summer of 1998 where this work has
been started.
\end{acknowledgements}

\end{document}